\documentclass[prb,twocolumn,aps,showpacs]{revtex4}
\usepackage{graphicx}
\newcommand{\beq}{\begin{eqnarray}}
\newcommand{\eeq}{\end{eqnarray}}

\begin{document}
\title{ Experimental Detection of Sign-Reversal Pairing in Iron-Based Superconductors }
\author{Jiansheng Wu and Philip Phillips}
\affiliation{Department of Physics, University of Illinois at
Urbana-Champaign, 1110 W. Green Street, Urbana IL 61801, U.S.A.}

\begin{abstract}
  We propose a modified Josephson corner-junction experiment which
  can test whether the order parameter in the iron pnictides changes sign between the electron and hole pockets of the Fermi surface.
\end{abstract}


\maketitle

\begin{figure}
   \includegraphics[bb=133 318 709 817,width=2in]{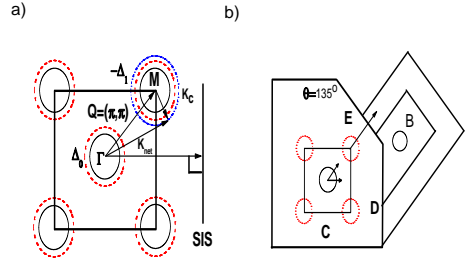}
   \caption{Design of the SQUID junction to test $s_{\pm}$-wave superconductivity. a) Fermi surface in the folded Brillouin zone to show
   the principle of design. b) The SQUID design. The left-hand-side of the junction is an iron-based SC which is cut on the $[010]$ (E face),
    $[110]$ (D face) and $[1\bar{1}0]$ (C face) planes. Any two of the planes are connected to a conventional
   $s$-wave SC on the right-hand-side through standard oxide-barrier SIS junctions\cite{Dale}. The magnetic field is perpendicular
   to the plane.}\label{SQUID}
   \end{figure}

Iron pnictides represent the newest member of the class of
correlated materials in which superconductivity emerges from doping
an ordered state\cite{Wang,Cruze}. One of the intriguing proposals
for superconductivity in these muti-band systems  is that spin
fluctuations mediate electron pairing between different regions of
the Fermi surface but with different signs for the order
parameter\cite{s}. In the unfolded Brillouin zone, the regions of
the Fermi surface which are relevant are the electron and hole
pockets located at the $M$ and $\Gamma$ points\cite{Liu},
respectively as illustrated in Fig. (\ref{SQUID}a). The result is a
nodeless gap, denoted as $s_\pm$, with a rough momentum dependence
of $\cos k_x\cos k_y$\cite{s}. While the preponderance of the
experiments support isotropic nodeless
superconductivity\cite{Hashimoto,Mu,Liu} in both the 1111 and 122
materials, the power-law behaviour of the spin-lattice relaxation
rate, $T_1^{-1}\approx T^3$ has been used as a strong indication of
line nodes\cite{Nakai2}.  However, nodeless $s_\pm$ pairing has
recently been shown to also yield $T^3$ behaviour at high
temperatures\cite{sNMR}. Fine tuning with disorder is necessary to
obtain the $T^3$ dependence of $1/T_1$ at low temperatures.
Alternatively, superconductivity with multiple gaps can also give
rise to such a deviation from the standard BCS exponential fall-off
of $T_1^{-1}$. In fact, in superconductivity in
Ba$_{0.6}$K$_{0.4}$Fe$_2$As$_2$ and PrFeAsO$_{0.89}$F$_{0.11}$ is
consistent with at least two gaps with ratios of $2$ and $3.2$,
respectively.  Hence, the complete consistency of $s_{\pm}$ pairing
with the experimental data is far from settled.


Nonetheless, given the novelty of the $s_{\pm}$ state, it is
important to definitively determine its relevance to the pnictides.
Although there are some proposals on phase sensitive measurements
such as a three layer sandwich structure\cite{sandwich}, they are not
direct probes of the order parameter phase.
As phase-sensitive measurements\cite{Dale} using Josephson interferometry were
pivotal in settling the question of the symmetry of the order
parameter in the cuprates, we focus here on whether or not such a
similar experiment can be performed to falsify the claim that the
order parameter in the pnictides has $s_\pm$ symmetry. Detecting an
$s_{\pm}$ state in the pnictides poses a distinct challenge from
discerning the sign change in the $d_{\rm x^2-y^2}$ state in the
cuprates because the sign change occurs along the crystal
axes\cite{Dale}.

\begin{figure}
   \includegraphics[bb=133 318 709 817,width=2.5in]{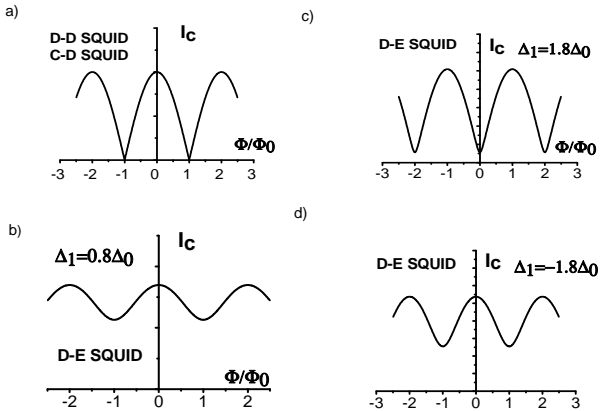}
   \caption{All possible interference pattern for different $x$ values where $x$ is defined as $\Delta_1=-x\Delta_0$. a) For junction connecting D-D or C-D faces;
    b) c) d) are all junction connecting D-E faces. b) $x\in[0,1]$; c) $x\in[1,\infty)$; d) $x\in(-\infty,0]$.
   }\label{Junction}
   \end{figure}

Central to the design of any standard
superconductor-insulator-superconductor (SIS) junction oxide-barrier
is the highly directional nature of the transport.  Namely the
junction can only detect the order parameter in the direction
perpendicular to the crystal face\cite{Dale}.  It is for this reason
that a standard corner junction can be used to detect the sign
change of the d$_{x^2-y^2}$ order parameter because the order
parameter has a natural alignment along the crystal axes.  In this
sense, detecting the s$\pm$ state depicted in Fig. (\ref{SQUID}b)
poses a distinctly new challenge because no such alignment of the
order parameter and the crystal axes is present.  Consider an
$s_{\pm}$ SC and a conventional $s$-wave SC joined by a weak link.
Let $\Delta_0$ and $\Delta_1$ be the magnitude of the order
parameters at the $M$ and $\Gamma$ points, respectively in the
iron-based superconductor.   The SQUID design is shown in
Fig(\ref{SQUID}b).  Let us consider the gap on the $M$-pocket which
experimentally is less than $\Delta_1=25\rm meV$\cite{Ding}. A gap
of this magnitude corresponds to a wave vector for the
center-of-mass of a Cooper pair  emanating from the
  $M$-pocket that is less than
$K_c=m_{\rm eff}\Delta_1/\hbar p_F\approx
\frac{0.026}{k_F}\left(\frac{m_{\rm
eff}}{m_e}\right)\left(\frac{\pi}{a_0}\right)^2$ where $m_{\rm
eff}$,$m_{\rm e}$,$a_0\approx 2.83\dot{A}$,$p_F$ and $k_F$ are the
effective mass,electron mass, lattice constant, Fermi momentum and
Fermi wave vector respectively. This number is much smaller than the
wave vector $Q=(\pi,\pi)/a_0$ , in the folded Brillouin Zone, by a
factor of $3$.
 In this case, regardless of the direction of the
Cooper pairs, it is impossible to choose the wave vector ${\bf
K}_{\rm c}$ of a Cooper pair emanating from $M$ so that the net wave
vector ${\bf K}_{\rm net}$ is perpendicular to the C or D faces of
the junction in Fig. (\ref{SQUID}b). However, this condition is
easily met at the
 face E which lies at an angle of $45^o$ from the horizontal.
 Such scattering of a Cooper pair from $\Gamma$ to the $M$ point requires Umklapp scattering as the net momentum transfer is $\bf Q$.
By contrast, the order parameter associated with $\Gamma$-pocket in
the folded zone can be sensed by all faces. Taking this into
account, we compute the associated critical current for a SQUID
joining the surfaces C and D, D and D, or D and E.  For the former
two, the critical current is given by $2\Delta_0\sin(\Phi/\Phi_0)$
where $\Phi$,$\Phi_0$ is the magnetic flux and flux quantum. This is
the standard s-wave result.  However, for a D-E SQUID, the situation
is different; the critical current,
\begin{equation}
\sqrt{\Delta_0^2+(\Delta_0-\Delta_1)^2+2\Delta_0(\Delta_0-\Delta_1)\cos(\Phi/\Phi_0)}
\end{equation}
is governed by the magnitude and sign of the order parameter at the
$M$ and $\Gamma$ points as depicted in Fig. (\ref{Junction}).  To
recover the standard s-wave result simply requires reversing the
sign of $\Delta_1$. All the possible interference patterns as a
function of $\Phi/\Phi_0$ are shown in Fig.(\ref{Junction}). For the
$1111$ pnictide material, there are two hole pockets at the $\Gamma$
point with $\Delta_0=(6+12) \rm meV$ and two electron pockets at
$\rm M$ point with $\Delta_1=(12+12)\rm meV$.  This case corresponds
to (c) in Fig. \ref{Junction})\cite{Ding}.  As all the possible
interference patterns differ substantially from the standard s-wave
result, this experimental design should offer a definitive test of
$s_\pm$ pairing in the pnictides.

{\bf Acknowledgement } J. Wu would like to thank Anthony J. Leggett
, Dale J. Van Harlingen, B. Andrei Bernevig and  Seungming Hong for
their helpful discussions.

{\bf Note} After this paper was completed, a similar idea
was proposed by I. I. Mazin and D. Park in arXiv:0812.4416. We offer
here 
a more complete explanation why along the $D$-face, only holes contribute
the current while along the $E$-face, both electrons and holes do.
Further, $D$-$E$ have a $\pi$ phase shift only when the gap on
electron Fermi surface is larger than the gap on hole Fermi surface
which is not mentioned in arXiv:0812.4416.

\end{document}